\documentclass[a4paper,11pt,twoside]{article}
\usepackage{a4wide}

\newcommand{\eps}{{\epsilon}}

\newcommand{\beq}{\begin{equation}}
\newcommand{\eeq}{\end{equation}}
\newcommand{\beqa}{\begin{eqnarray}}
\newcommand{\eeqa}{\end{eqnarray}}
\newcommand{\ben}{\begin{enumerate}}
\newcommand{\een}{\end{enumerate}}
\newcommand{\bi}{\begin{itemize}}
\newcommand{\ei}{\end{itemize}}

 \title{Interpreting Quantum Chromodynamics from Spin Glass- and Polymer
 Analogies}

\author{U.\ Krey\\
Inst.\ f\"ur Physik II, Universit\"at Regensburg, 93040 Regensburg,
Germany } 

\date{Feb. 11, 2010}

\begin{document}

\maketitle
\begin{abstract}

\noindent In an informal way some kind of Ising Lattice QCD is
introduced which allows to interprete and discuss the well-known theory
of Quantum Chromodynamics (confinement, quarks and gluons etc.) from
simple phenomena of magnetism and polymer physics. Also the viewpoint of
duality is stressed. Moreover, the non-abelian character of the Lie
Group SU(3) is important. The essential arguments for confinement and
asymptotic freedom are given at the very end of the manuscript, after an
appendix.

\end{abstract}
{
\vglue 0.2 truecm\hrule\vglue 0.5 truecm

\section{Introduction}

 This is an informal paper, not intended for publication\footnote{e-mail
uwe.krey@physik.uni-regensburg.de,\newline $\,\quad$updated versions at
www.physik.uni-regensburg.de/forschunng/krey }. The QCD
(Quantum Chromodynamics. i.e. the theory of strong interaction) is
perhaps the most-important theory of present-day high energy physics,
dealing with quarks, gluons, confinement, and perhaps the quark-gluon
plasma transition predicted to happen at ultra-high energies around 200
GeV. In contrast, ferromagnetism has characteristic temperatures around
1000 K or below, corresponding to 0.1 eV, and spin glass magnetism
happens even below 30 K. Thus, since the phenomena differ by 12 or more
energy decades, one would tend to believe that they are unrelated.
However, one should be warned by the facts that (i) the seminal paper of
Franz Wegner on lattice-gauge theories, \cite{Wegner}, is based on Ising
spins, and (ii) that a strongly pedagogical and interdisciplnary paper
of John Kogut, \cite{Kogut}, also stresses the relation between QCD and
spins, and (iii) that according to a recent colloquium talk of Fritjof
Karsch, \cite{Karsch}, a 2nd-order phase transition to a quark-gluon
plasma, which seems to exist at temperatures around $ T_c\approx 200$
GeV/$ k_B$, belongs apparently to the universality class of the 3d Ising
model.

 \section{Spin glasses and "frustration" }

In fact, in the theory of Ising spin glasses one is working with the
Hamiltonian

\beq
{\cal{H}}=\,{\bf { -}}\, \sum_{i,k}\, s_i \,J_{i,k}\, s_k\,,\eeq

Here the spin variables are Ising  degrees of freedom, $s_i=\pm 1$, and the
$J_{i,k}$ are real numbers, e.g. quenched Gaussian random numbers, or they
are also binary random numbers. In the last-mentioned case, where one is
dealing with a so-called $\pm 1$ spin glass, the notion of "frustration"
plays an important role: One considers a closed loop ${\cal W}$ on the
lattice containing the sites i and k, and on this loop one considers the
product $P_W\,:=\,J_{i,k}\cdot J_{k,l}\cdot ...\cdot J_{n,m}\cdot J_{m,i}$.
The loop ${\cal W}$ is called "frustrated" , \cite{Toulouse}, if the product
is ${\it negative}$.

\section{Relation to QCD} The relation to the Quantum Chromodynamics is
essentially a matter of three steps: 

\bi 

\item At first on interpretes the spin variables $s_i$ as "quarks" and the
$J_{i.k}$ as "gluons" . (This is of course only semantics).

\item Secondly, {\bf both} quantities are now allowed to {\bf fluctuate}
according to the thermal laws. (This is an essential physical  step).

\item Finally one adds to ${\cal H}$ the so-called {\bf  Wilson loop}
contributions $ J_{i,k}J_{k,l}J_{l,m}J_{m,i} $, i.e. the so-called plaquette
variables, defined on the boundary of the side faces of the
four-dimensional elementary hypercubes defining the four-dimensional
euclidean lattice corresponding to the $(3 \,\,{\rm plus}\,\,
1)$-dimensional Minkowski space underlying at T=0 the quantum
chromodynamics. In this way one obtains some kind of "Ising Lattice QCD".

\ei

Besides: This contribution corresponds to the above-mentioned
"frustration", but one works no longer with a "quenched" situation, but with
"annealed" degrees of freedom.

\section{Gauge invariance}
The Hamiltonian is now explicitly gauge invariant, namely against the
simultaneous transformations

\beq
s_i\to s_i\cdot\eps_i ,\quad s_k\to s_k\eps_k ,\quad J_{i,k}\to
\eps_k\,J_{i,k}\,\eps_k\,.
\eeq

Here $\eps_i=\pm 1$ and $\eps_k=\pm 1$ are independent random numbers.

Of course this is gross simplification. In the QCD, the $\eps$
parameters are  replaced by SU(3) matrices. But the essential philosophy
is recovered by the Ising approximation.

This corresponds to a very large extra degeneracy, which already plagues the
spin glass physics at low temperatures and also comes into play in the QCD
(see below) and means that one is dealing with a very large entropy.

\section{Analogies and Consequences} The gauge invariance leads to the
notion of a Mattis spin glass, \cite{Mattis}, i.e. a configuration, looking
perfectly random, i.e. $s_i\equiv \eps_i$, but which is actually a
"ferromagnet in disguise".

This example means that, in any case, one should be very careful with
observations.

We start with parallel one-dimensional ferromagnetic Ising chains, e.g. all
in the direction of the time axis. If the "gluons" are quenched, as in the
spin-glass analogy, while the "quark" degrees of freedom are allowed to
fluctuate thermodynamically, then one has of course weak interaction at
large distances and strong interaction at neighbouring lattice points.  This
is just the ${\it opposite}\,$ situation as in Quantum Chromodynamics: there
one has ${\it asymptotic\,\,freedom}\,$ at low distances and ${\it
confinement}\,$ at large ones, \cite {Wilczek}.

In fact, the kink excitations in one-dimensional Ising model, see
\cite{Kogut}, in particular those on vertically parallel Ising chains,
immediately destroy the long-range order, which in the one-dimensional
Ising model exists only at T=0. As a consequence, right from the
beginning one should not work with quenched gluon degrees of freedom, as
in spin glasses but better with the dual situation of quenched quark
degrees of freedom. In fact this was the preferred approximation in the
early years of QCD, as sketched e.g. in the standard book of Becher,
B\"ohm and Joos, \cite{Joos}.

\section {Duality}

The importance of duality has already been stressed in the seminal paper
of F. Wegner, \cite{Wegner}, where it has been stated among other points
that the low temperature results of the original model corresponds to
the high temperature results of a dual one. To be specific, the frozen
gluon approximation of the spin-glass theory and the frozen quark
approximation of the early years of the QCD are somehow dual to each
other, and the asymptotic freedom as a short-distance behaviour of the
QCD is somehow dual to the usual behaviour of magnetic phases at long
distances.

Moreover, also the usual 1/r-behaviour of V(r) in the non-confined
phases and the strange r-behaviour of the stretched loops under
confinement can be considered as somehow {\it dual\,} to each other.

Here the caveat "somehow" should not be overlooked. In fact, the
original models considered by \cite{Wegner} are Ising models with a
gauge invariant 2n-point interaction in d dimensions, which is relevant
in our case   for n=2 and d=4 (respectively for d=3 at high
temperatures, where in statistical mechanics a phase-space integral
along the time axis is replaced by a sum of finite Matsubara
frequencies, see e.g. Abrikosov et al, \cite{Abrikosov}). The
Hamiltonian corresponds to the isolated gluon Hamiltonian

 $${\cal H}=K\,\sum_{\,\partial\{plaquettes\}}\,
 J_{i,k}\,J_{k,l}\,J_{l,m}\,J_{m,i}\,,$$

with positive K, taken around the edges of a "plaquette", i.e. over the boundary of all
faces of the d-dimensional simple cubes, defining our lattice. For d$\le
3$ this creates a simple d-dimensional hypercubic lattice itself, the
"dual" one, since e.g. $3\choose 2$ is 3 again. However for d=4,
$4\choose 2$ is 6, i.e. there are now too many faces for simple duality.

However, in this case one may distinguish the three faces going out from
"the lower left" of a hypercube from those coming in towards the "upper
right", i.e. one has the possibility to get different parity with respect to
the center of the hypercube, and may in this way distinguish "electric" and
"magnetic" field quantities (see below.)

 \section{Rubber elasticity}

On the other hand the above-mentioned crucial behaviour characterizing
the confinement, namely the linear increase of V(r) with increasing
distance r, which is always compared on pedagogical reasons  with the
behaviour of a stretched rubber band, has also a direct physical
meaning: it is well known that rubber elasticity means {\it entropy
elasticity}, and is physically related with the entangling,
disentangling and deforming of the  many loops in polymers,
\cite{Goeritz}. Thus one should not be astonished that here Wilson
loops, i.e. the dominance of gluon disorder ("annealed disorder", to be
precise!) leads to the phenomenon of confinement. Essentially it is the
huge degeneracy mentioned above, i.e. essentially a consequence of gauge
invariance which is finally responsible for the rubber-like confinement.

\section{Random fields}

If the quarks are frozen and the gluons fluctuating, one can describe the
first part of our Hamiltonian, the term usually called the "spin glass
Hamiltonian", i.e. (1), as a "random field term", where the random fields
for the gluons, i.e. for the link variables $J_{i,k}$, are given by the
momentaneous values of the variables $s_i$ and $s_k$. However, once again,
one has not "quenched" randomness, but annealed one, and the gauge
invariance should be important again.

\section{Polymer States and Confinement}

There is another virtue of the polymer analogy: the fact that in these
systems there is apparently {\it everywhere\,} a finite probability to have
entanglement between different loops makes it plausible that in these
systems one would have an {\it area law\,} behaviour, and not a {\it
circumferential\,} one, of the Wilson loop variables. As is known, this is
the prerequsite for confinement.

Here some remarks about the difference between QED (Quantum
Electrodynamics) and the QCD (Quantum Chromodynamics) are in order; In
the QED, with an abelian gauge group, the U(1), the loop segments are
uncharged, i.e. unstructured, whereas in the QCD (non-abelian gauge
group SU(3) ) the loop segments carry colour-charge, i.e. the loop
segments themselves generate a colour-electric field "$\bf E$" or a
colour-magnetic field "$\bf B$" (quantities with 8 complex
3x3-matrices);  i.e. the segments themselves generate the
surface-densities "$\rm{rot\,\,}\bf E$" or "$\rm{rot\,\,}\bf B$" giving
rise to the area-law behaviour. These topological consequences of the
non-abelian character of the gauge group, which creates some kind of
entangling property of the loops, corresponds in polymer physics to the
presence of chemical units, by which loop segments are structured, such
that different loops are linked together.

Concerning the non-abelian character of the Wilson products $P_W
=J_1\cdot\, ...\,\cdot J_4$ (using a symbolical writing) in the groups
SU(3),  the following remark is in order: Since one is working with a
closed surface (i.e. the sum of the edges along the faces of  a simple
hypercube, the sum of the contributions from the faces can be
transformed into a volume integral by a kind of Gaussian theorem.
However, using general statements similar to "div rot $\equiv$ 0" one
gets zero for the abelian case. In contrast, if the J-quantities are
taken from a non-abelian Lie group, then the above-mentioned identity
must be replaced by a nontrivial one involving the structure constants
of the group.

\section{The quark-gluon plasma}

Ultimately, e.g. if temperatures or pressure correspond roughly to the
mass of the heaviest quark, 170.9 GeV, one should definitely give up any
quenched approximation and should treat quarks and gluons on the same
footing, namely both as equally fluctuating quantities. The
corresponding behaviour leads to the prediction of a 2nd order
quark-gluon plasma phase transition, which corresponds to the
universality class of the 3d Ising model. That one has a
three-dimensional lattice, and no longer a four-dimensional one, is
natural, since in the numerical simulation the time-axis is no longer
critical, \cite{Karsch}. Moreover, already Franz Wegner, with his
generalized Ising ansatz, had a precise duality relation leading to the
3d-Ising model for this case.

\section{Conclusion}

The phenomena of lattice QCD , i.e.  Quantum Chromodynamics, quarks,
gluons, confinement, gauge-invariance, the quark-gluon plasma
transition, and so on, have been given certain pedagogical plausibility
interpretations by referring to well-kown phenomena in solid-state
physics, e.g. spin-glass and polymer behaviour, in particular to a
"Mattis spin glass", i.e. a "ferromagnet in disguise", \cite{Mattis},
and to random-field physics, entanglement-probability, and "frustation".

Moreover, the property of {\it duality\,} is stressed: i.e. the
high-temperature phase of the {\it quarks} (or at least dominated by
them) seems to correspond by duality to the low-temperature phase of the
{\it gluons} (or dominated by them). At the same time, by this duality,
there is apparently a 1/r $\to $ r correspondence.

The duality is a direct one, since the centers of the simple hypercubes
of the original lattice correspond to the sites of the dual lattice.
Also the splitting of the colour-fields into electric and magnetic parts
can be understood on topological reasonings, and the importance of the
non-abelian group character is seen.

Then, finally, at temperatures of roughly 160 GeV/$k_B$ the quark-gluon
plasma transition happens, where both particles should be treated on
equal footing and where, on thermal averaging, the non-commutativity is
lost, and also the criticality along the time-axis, such that the
transition is of the simple universality class of the 3d-Ising model.

\section*{Appendix and some arguments}

The term corresponding to "div rot $P_W$" or a non-abelian Lie group is
well known to be

\beq \rm{"div\,\, rot\,\, P_W"}\to f^{a\,,b}_{\,c}\,(P_W)^{\nu
,\,\mu}_b\,(P_W)^c_{\mu ,\,\nu}\,,\eeq where as usual one has to sum
with respect to the indexes appearing two times. Th\,e
$\rm{f^a_{\,b,c}}$-quantities are the so-called structure constants of
the Lie group, antisymmetric with respect to b and c. The indexes a,b,
and c are shifted from upwards to downwards and vice versa with the
trivial signature (+,...,+) from a=1 to a=8. Also the $\mu$ and $\nu$
indexes are transformed with the trivial signature, but a
four-dimensional one, since one is working in Euclidean four-space. This
a-dependent quantity (a=1, ... , 8) has to be discussed on the "dual"
lattice, i.e. on the the centers of the hypercubes.

Now the important qualitative arguments: \bi

\item At high energies, on the one hand, this "pure gluon" term, which
apparently leads to $confinement$, is most important, $\propto E^8$,
compared with that one where the quark-gluon coupling appears, i.e.
compared with the "spin glass" one, $\propto E^1$.

\item On the other hand, the {\it structure constants\,}
$\rm{f^{a\,,b}_{\,c}}$ keep their ${\cal O}(1)$ values, i.e. the
non-abelian character becomes more and more neglegible with increasing
energy even at relatively low temperatures. This is the {\it asymptotic
freedom}.

\item At very high temperatures the non-abelian character is neglegible
also at smaller energies, and - as stressed above - there is the
possibilty of a continuous phase transition to a quark-gluon plasma,
with the universality class of the 3d-Ising model. \ei

}
\end{document}